# Commercial AlGaAs diodes provide simplified feedback for adaptive femtosecond pulse compression

PATRYCJA H. WYSOWSKA, ADRIAN CHLEBOWSKI, ŁUKASZ A. STERCZEWSKI, AND JAKUB BOGUSŁAWSKI*

*Laser & Fiber Electronics Group, Faculty of Electronics, Photonics and Microsystems, Wrocław University of Science and Technology, Wybrzeże Wyspiańskiego 27, 50-370 Wrocław, Poland*
**jakub.bogusławski@pwr.edu.pl*

**Abstract:** Maintaining high peak power in ultrashort laser pulses is essential for efficient nonlinear interactions, particularly in multiphoton microscopy. Dispersion introduced by optical components broadens these pulses and degrades their peak power, thus motivating adaptive pulse compression guided by pulse-width-dependent feedback. However, the conventional second-harmonic generation feedback requires a nonlinear crystal, spectral filtering, and careful optical alignment. Here, we investigate commercially available AlGaAs laser and light-emitting diodes as alternative two-photon absorption feedback sensors. Their photocurrent exhibits a quadratic dependence on average optical power and inverse dependence on pulse duration, which is consistent with the analytical model predicting maximum response for the shortest, single-peaked pulse. Both devices generated a measurable photocurrent at sub-milliwatt average power levels and showed a weaker polarization dependence than second-harmonic-generation-based solution. All three feedback methods achieved comparable mean compressed pulse durations of 210–228 fs. These results demonstrate effective adaptive pulse compression with a substantially simpler feedback.

## 1. Introduction

Ultrashort laser pulses play an essential role in numerous scientific and technological applications, particularly those relying on nonlinear optical interactions [1]. However, propagation through optical components, such as lenses and optical fibers, inevitably introduces dispersion that broadens pulses in time and reduces their peak power [2, 3]. A prominent example is multiphoton microscopy – an important technique in biomedical imaging – where nonlinear excitation strongly depends on the instantaneous optical intensity [4]. Consequently, dispersion-induced pulse broadening reduces the fluorescence signal at a given average optical power, which degrades image quality [5, 6]. Various pulse compression techniques, including prism compressors, grating compressors, and chirped mirrors, have been developed to compensate for this broadening [7]. Although widely used, these methods offer limited flexibility in compensating for complex spectral phase distortions, particularly those involving higher-order dispersion [8, 9]. Furthermore, these methods often require manual adjustments to compensate for changes in dispersion, limiting their practicality in applications requiring frequent recalibration. In contrast, adaptive pulse compression provides an alternative approach by employing a spatial light modulator (SLM) to control the spectral phase of ultrashort pulses [9–11]. Combined with an optimization algorithm, automated compensation of phase distortions is possible without requiring prior knowledge of the spectral phase [12–14].

In general, adaptive pulse compression requires a feedback signal sensitive to changes in pulse duration, which serves as the objective function for the optimization algorithm [15]. Second-harmonic generation (SHG) is widely used as a nonlinear feedback mechanism in automated optimization of ultrashort pulses. Depending on the optimization objective, either the integrated SHG intensity or its spectrum can be used to evaluate and optimize the output pulse characteristics [9, 16, 17]. However, SHG-based detection requires a nonlinear crystal, spectral filtering of unconverted fundamental wavelength, and careful optical alignment due to phase matching requirement, increasing the complexity of the experimental setup [6, 18]. Consequently, alternative feedback mechanisms that provide comparable sensitivity to pulse duration while simplifying the detection arrangement are of considerable interest.

One such alternative relies on two-photon absorption (TPA) in semiconductor devices [19]. This third-order ($\chi^{(3)}$) nonlinear process occurs when two photons are absorbed simultaneously, allowing charge carriers to be excited across the semiconductor bandgap when energy of a single photon is insufficient. The resulting photocurrent is proportional to the square of the optical intensity, providing a signal sensitive to changes in pulse duration. Commercial semiconductor devices have already been successfully employed for ultrashort pulse characterization [20]. For instance, Siegner et al. demonstrated that TPA-based feedback from a GaAsP photodiode can also enable adaptive pulse compression, yielding spectral phase correction comparable to that obtained using SHG [19]. More recently, Fabry-Pérot laser diodes (LDs) have shown promising performance as sensitive detectors of femtosecond pulses [21]. However, their waveguide geometry can make detected signal sensitive to optical coupling condition and polarization [21]. Light-emitting diodes (LEDs) offer an alternative device geometry and have also been demonstrated as TPA detectors, including AlGaAs LEDs operating at wavelengths around 1 μm [22, 23]. Despite this, their potential as feedback detectors for adaptive pulse compression remains largely unexplored. To best of our knowledge, adaptive pulse compression using commercially available AlGaAs LDs or LEDs as feedback detectors, together with a systematic comparison against SHG, has not been demonstrated yet.

Here, we systematically evaluate the performance of commercially available AlGaAs LDs and LEDs as feedback detectors for adaptive femtosecond pulse compression. Their nonlinear responses are characterized in terms of optical power, pulse duration, and polarization dependence, using conventional SHG detection as a reference. We then employ each detector in an SLM-based adaptive compression system, comparing the achieved pulse durations and optimization repeatability under two different initial dispersion conditions. The experimental investigation is further supported by a theoretical analysis of the nonlinear photocurrent response, demonstrating how its maximization can be used to guide pulse optimization. The results demonstrate that semiconductor-based feedback enables adaptive pulse compression with performance comparable to conventional SHG-based feedback, offering a simpler alternative for automated dispersion compensation in nonlinear optical applications, particularly multiphoton microscopy.

## 2. Methods

### *2.1. AlGaAs semiconductor devices*

Two commercially available AlGaAs devices were investigated: a Fabry-Pérot LD with a nominal emission wavelength of 920 nm and LED with a nominal emission wavelength of 940 nm. Both devices are widely used in industrial applications, offering high reproducibility and stable, well-defined performance. The diodes were illuminated with femtosecond laser pulses centered at 1032 nm and operated in zero-bias mode to minimize the contribution of dark current related noise. The generated photocurrent was converted into voltage signal using a transimpedance amplifier (TIA, FEMTO DHPCA-100). The electrical signal was processed using a low-pass filter (LPF, 10 kHz) to suppress high-frequency noise during both detector characterization and adaptive pulse compression.

### *2.2. Experimental setup*

The experimental setup is shown in Fig. 1. The home-built Yb:fiber laser emits highly-chirped pulses with a duration of 14.8 ps, centered at 1032 nm, with a spectral bandwidth of 11 nm and a repetition rate of 30 MHz. The laser spectrum presented in Fig. 1 was measured using a Broadcom Qmini spectrometer. The pulses were initially compressed using a transmission-grating compressor, which also allowed the residual dispersion to be adjusted by varying the optical path length using a diffraction grating mounted on a linear translation stage. The pulse duration at the compressor output could be adjusted from 220 fs to 1.6 ps, with an average output power of approximately 35 mW. The compressor was followed by a half-wave plate (HWP) to adjust the beam polarization for the subsequent pulse-shaping stage.

For adaptive pulse compression, the spectral phase was controlled using a liquid-crystal spatial light modulator (LC-SLM, Jenoptic SLM-s640). The device features 640 independently addressable pixels and provides a phase modulation up to 7π. The LC-SLM was incorporated into a folded 4f pulse shaper consisting of a reflective diffraction grating (DG, 1400 lines/mm, Spectrogon) and concave mirrors (CM) with a focal length of 500 mm, resulting in a spectral resolution of approximately 0.1 nm per pixel.

The modulated beam was then split using a 50/50 ultrafast beam splitter (BS, Thorlabs UFBS5050). One arm was directed to an autocorrelator (APE pulseCheck), allowing the pulse duration to be monitored during optimization. The other arm was directed to the feedback detection module, where generated signal was measured and used by the optimization algorithm to evaluate successive SLM phase masks.

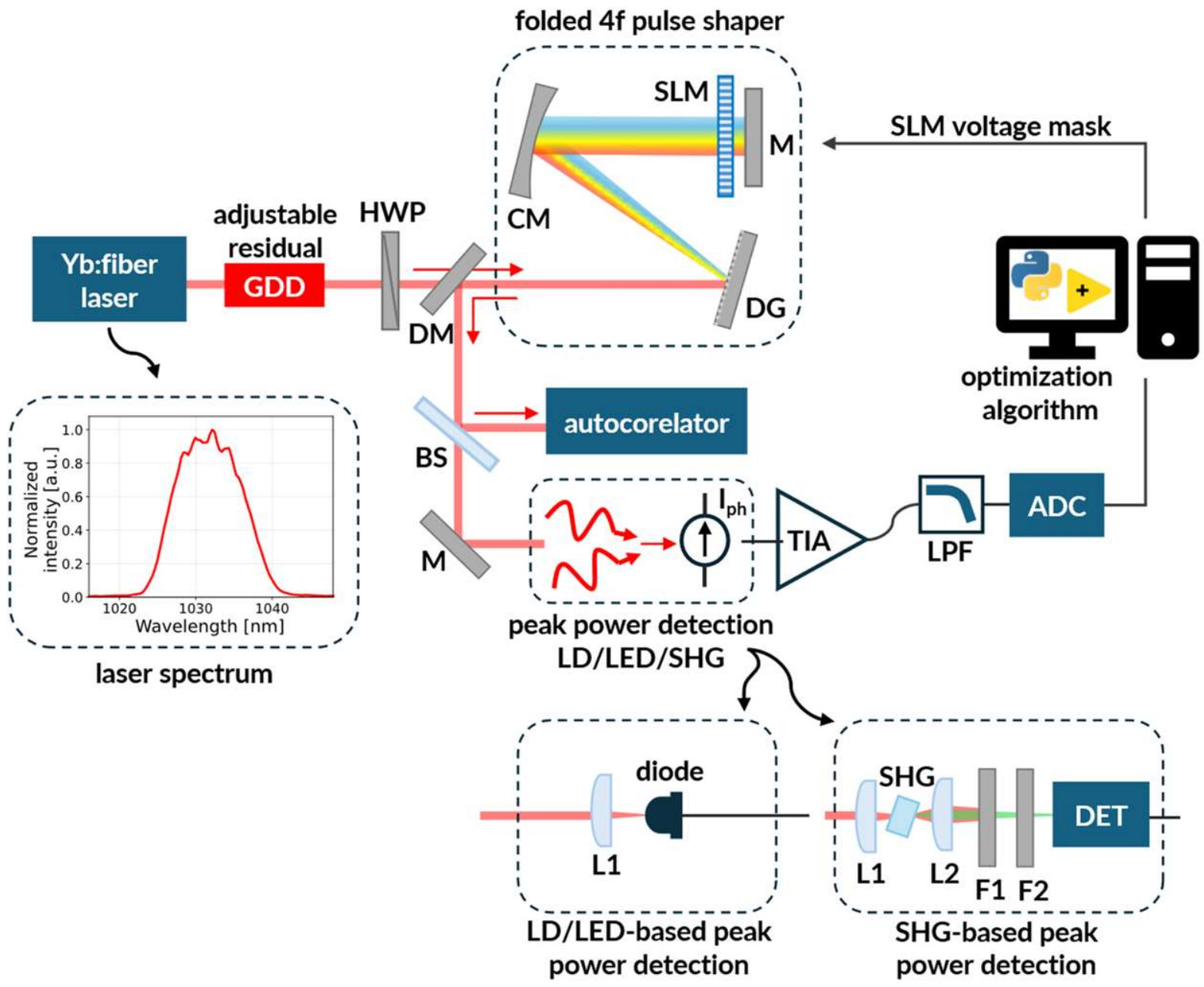

Fig. 1. **Experimental setup for adaptive pulse compression.** Pulses from a Yb laser (spectrum shown in the inset) were initially compressed using a transmission-grating compressor with adjustable residual group-delay dispersion (GDD). A folded 4f pulse shaper equipped with an LC-SLM was used for adaptive compression. The beam was split between an autocorrelator and an interchangeable feedback module employing an AlGaAs LD, an AlGaAs LED, or an SHG-based reference peak power detection. The detected signal was amplified (TIA), low-pass filtered (LPF), and analog-to-digital converter (ADC) to provide feedback for the optimization algorithm controlling the SLM voltage mask. HWP – half-wave plate; M – mirror; DM – D-shaped mirror; BS – beam splitter; CM – concave mirror; DG – diffraction grating; L1, L2 – lenses; F1, F2 – optical filters; DET – photodetector.

Three feedback signals were investigated: those generated by the AlGaAs LD and LED, and a reference signal based on SHG. For semiconductor-based detection, each diode was mounted in a housing attached to an XYZ translation stage, allowing precise positioning relative to the incident beam. A focusing lens (L1, $f$ = 8 mm, Thorlabs A240TM-B) was used to focus the beam onto the active region of each diode. For a Gaussian beam with a diameter of approximately 2 mm at the L1, the calculated focal spot diameter was 5.3 µm at 1032 nm.

The semiconductor-based approach was benchmarked against a conventional SHG detection setup employing a 30 µm thick β-BBO crystal. The fundamental beam was focused on a crystal using an L1 to enhance nonlinear conversion efficiency. The generated second-harmonic light was then collected a second lens (L2, $f$ = 19 mm, Thorlabs AC127-019-C) and focused onto detector (DET, Thorlabs DET10A/1M). To suppress the residual fundamental radiation and isolate the SHG signal, a short-pass optical filter (F1, cut-off at 700 nm, Thorlabs FESH0700) and a bandpass filter centered at 510 nm (F2, FWHM = 42 nm, Thorlabs MF510-42) were placed in front of the detector.

In each approach, the generated photocurrent was converted into a voltage signal using TIA. The signal was subsequently low-pass filtered and digitized using the analog-to-digital converter integrated into the SLM control electronics. The acquired signal was then used as feedback for the optimization algorithm.

### *2.3. Optimization procedure*

Adaptive pulse compression was performed using a closed-loop optimization scheme based on a covariance matrix adaptation evolution strategy (CMA-ES) algorithm [24]. This derivative-free method was chosen for its ability to optimize an experimentally measured signal without requiring a mathematical model of the system or gradient information [25, 26]. The algorithm was implemented in Python and integrated with LabVIEW, which controlled LC-SLM and acquired feedback signal. Each CMA-ES generation involved evaluating 11 candidate phase masks. To account for the finite response time of the liquid crystals, a 200 ms settling delay was introduced after each mask update, followed by the acquisition and averaging of 200 ADC readings, which required approximately another 200 ms. Including SLM updates, data transfer, and algorithm computations, each generation took approximately 6 s.

To evaluate compression performance and repeatability, two initial dispersion conditions were investigated. The first involved approximately 44,000 fs$^2$ of GDD, resulting in a pulse duration of 600 fs, and was chosen to evaluate the ability of the feedback detectors to support compensation of substantial dispersion. The second condition involved approximately 11,000 fs$^2$ of GDD, resulting in a pulse duration of 260 fs. This lower dispersion was selected to assess the sensitivity of the feedback signals to modest pulse broadening, representative of dispersion introduced by microscope optics [4, 27]. Four independent optimization runs were performed for each feedback method (LD, LED, and SHG) under both conditions. The resulting pulse durations were measured using an autocorrelator.

## 3. Results

### *3.1 Comparative characteristics of LD, LED, and SHG responses*

The nonlinear responses of the investigated semiconductor devices were first characterized using an experimental setup consisting of a transmission-grating compressor, neutral density (ND) filter and focusing lens (L1), followed by the investigated detection pathways. The average optical power was controlled using the ND filter and measured with a power meter (Thorlabs PM400) with a silicon photodiode power sensor (Thorlabs S121C). The detector signal was amplified using a TIA, passed through a LPF and recorded with a Keysight InfiniiVision DSO-X 3034A digital oscilloscope.

Initially, the LD and LED photocurrents were measured as a function of average optical power at a fixed pulse duration of 220 fs (Fig. 2(a)). Measurements were performed at different TIA gain settings to avoid amplifier saturation at high optical powers and ensure a sufficient signal-to-noise ratio at low powers. The photocurrent was calculated from the measured voltage using the corresponding TIA gain. The measurements covered average optical powers ranging from approximately 30 µW to 30 mW. Both devices exhibited a quadratic dependence on average optical power at low powers, consistent with two-photon absorption, with measurable photocurrents obtained even at powers of a few tens of microwatts. In this regime, the LD generated a ~3.5 times higher photocurrent than the LED at an average optical power of 1 mW, potentially due to its waveguide and the resulting differences in optical coupling and interaction length. At even higher powers, i.e. above ~10 mW, the LD response began to saturate, whereas the LED continued to exhibit a quadratic response, with no signs of saturation observed over the measured range.

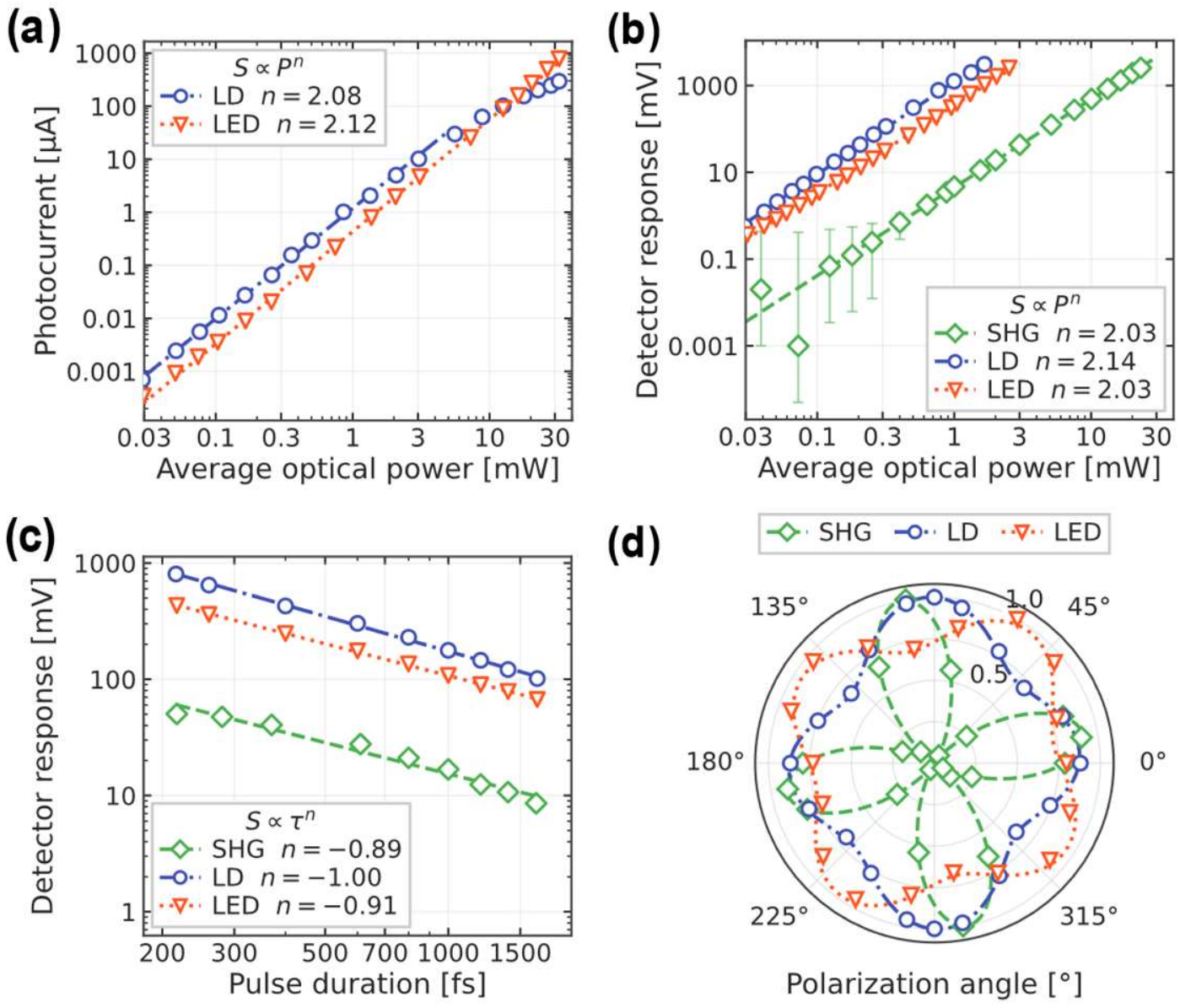

Fig. 2. **Comparison of the LD and LED detector responses with the SHG reference signal.** (a) Photocurrent generated by the LD and LED as a function of average optical power. (b) Nonlinear responses of the LD, LED, and SHG as a function of average optical power at a fixed pulse duration of 220 fs. (c) Nonlinear responses as a function of pulse duration at a fixed average optical power of 1 mW. (d) Normalized responses as a function of a linear polarization angle.

To further assess their suitability as peak-power-sensitive feedback detectors, the LD and LED responses were compared with a conventional SHG signal. First, the average optical power was varied using the ND filter while maintaining a fixed pulse duration of 220 fs (Fig. 2(b)). All three signals were recorded using a TIA gain of $10^6$ V/A. As shown in Fig. 2(b), all exhibited an approximately quadratic dependence on average optical power, with fitted exponents of 2.03, 2.14, and 2.03 for SHG, LD, and LED, respectively.

Next, the pulse duration was varied at a fixed average optical power of 1 mW (Fig. 2(c)). This was achieved by detuning the transmission-grating compressor to introduce different amounts of residual dispersion, with the resulting pulse durations verified using an autocorrelator. For these measurements, TIA gains of $10^6$ V/A and $10^7$ V/A were used for the semiconductor detectors and SHG signal, respectively, to improve the SHG signal-to-noise ratio. All three signals increased as the pulse duration decreased, following an approximately inverse relationship with fitted exponents of −0.89, −1.00, and −0.91 for SHG, LD, and LED, respectively. These results confirm that both semiconductor devices exhibit the expected nonlinear response and provide feedback sensitive to pulse duration, comparable in its power-law behavior to SHG.

To investigate the polarization dependence of the LD and LED and compare it with the SHG reference, an additional HWP was placed before the focusing lens L1. The HWP was rotated to vary the incident linear polarization, and the measured signals were normalized to their respective maxima (Fig. 2(d)). The SHG signal exhibited strong polarization dependence, with its amplitude decreasing to approximately 5% of its maximum value. In contrast, the LD and LED responses showed considerably smaller variations, with minimum values of 64% and 67% of their respective maxima. These results indicate that both semiconductor detectors are less sensitive to changes in incident polarization than SHG reference.

Overall, these measurements demonstrate that both LD and LED exhibit nonlinear responses to optical power and pulse duration, while showing considerably weaker polarization dependence than the SHG reference.

### *3.2 Adaptive pulse compression using LD, LED, and SHG feedback*

To evaluate their performance in adaptive pulse compression, the LD and LED were used as feedback detectors and compared with conventional SHG-based feedback under two different initial dispersion conditions. The resulting autocorrelation traces and optimized SLM voltage masks are presented in Fig. 3.

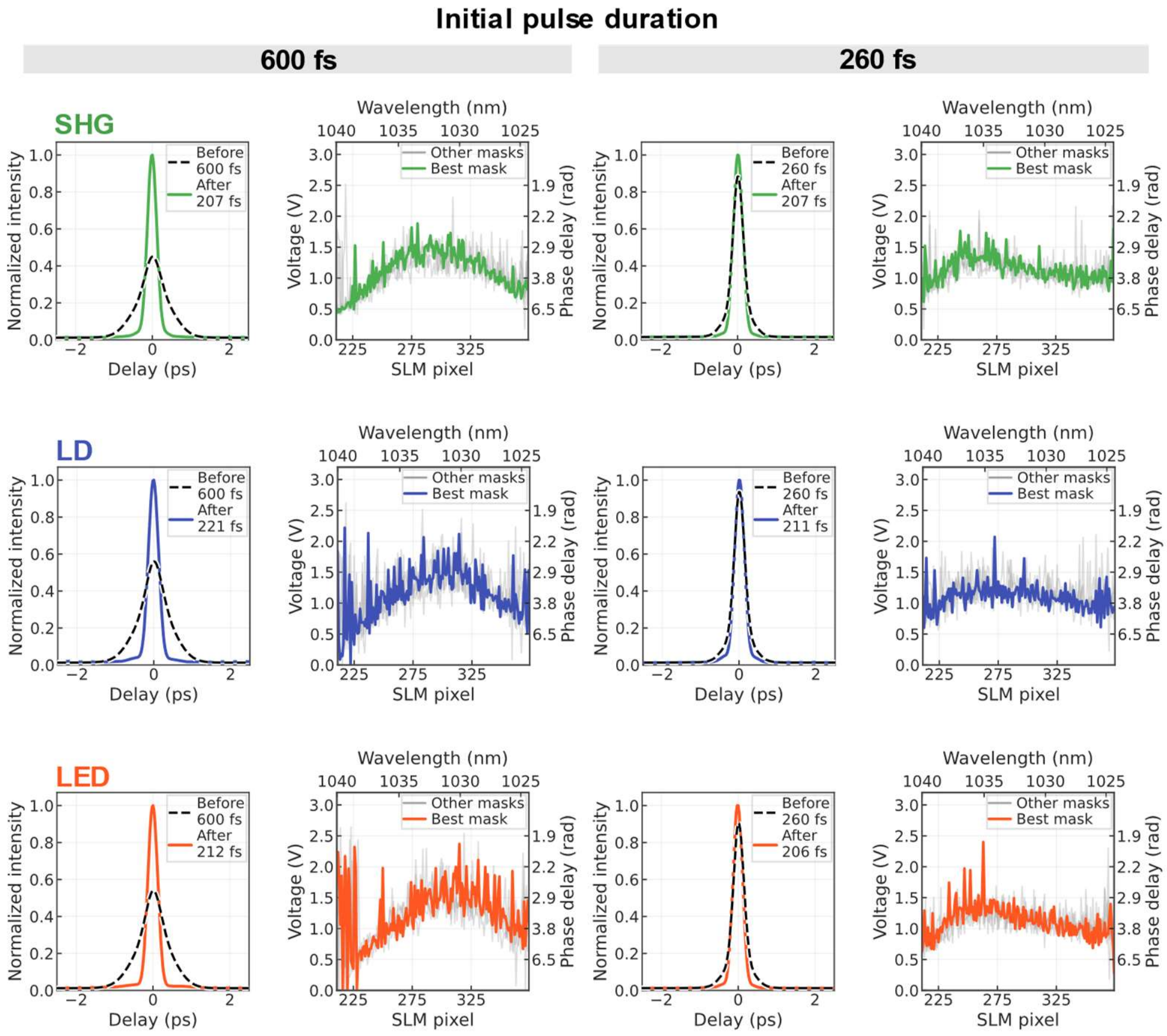

Fig. 3. **Comparison of adaptive pulse compression using SHG, LD, and LED feedback.** Rows correspond to SHG (top), LD (middle), and LED (bottom), with initial pulse durations of 600 fs (left pair) and 260 fs (right pair). For each condition, autocorrelation traces before (black dashed) and after (colored solid) optimization are shown alongside the corresponding SLM voltage masks. Both traces are normalized to the maximum of the compressed-pulse trace. Four independent optimization runs were performed for each condition, with the voltage mask producing the shortest pulse highlighted in color and the remaining three shown in gray. The colored autocorrelation trace corresponds to the highlighted mask. The SLM voltage, corresponding phase delay (right axis), and wavelength (top axis) are also indicated.

All three methods successfully compressed the laser pulses under both initial dispersion conditions. For pulses initially broadened to 600 fs, optimization significantly reduced the pulse duration, with both semiconductor detectors achieving compression results comparable to those obtained using SHG feedback. For initially 260 fs pulses, the initial pulse broadening was lower, yet all three feedback methods were still able to improve its temporal compression. While individual optimization runs produced different SLM voltage masks, their overall profiles remained similar under the same initial dispersion conditions, despite local variations.

The results of four independent optimization runs for each feedback method and initial dispersion condition are summarized in Fig. 4, including the final pulse durations, feedback gains, and generations at which the maximum feedback signals were recorded. For pulses initially broadened to 600 fs, the mean compressed pulse durations were 217.8 ± 7.3 fs for SHG, 228.0 ± 8.1 fs for LD, and 216.5 ± 3.9 fs for LED feedback. For initially 260 fs pulses, the corresponding values were 210.0 ± 2.6 fs, 215.8 ± 5.1 fs, and 210.0 ± 2.7 fs, respectively. The LED achieved mean compressed pulse durations comparable to those obtained with SHG under both initial conditions, whereas the LD resulted in longer mean pulse durations and larger standard deviations.

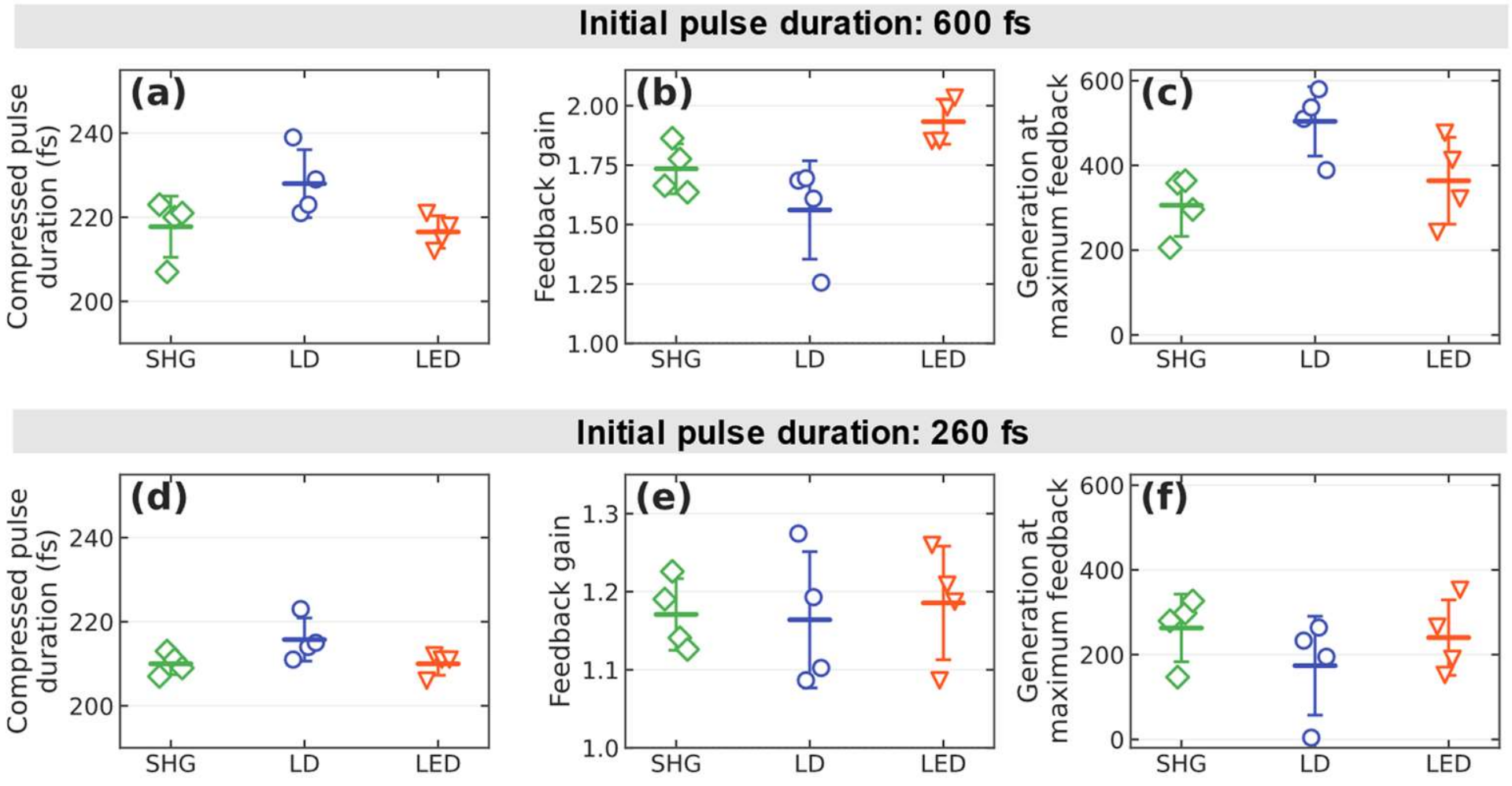

Fig. 4. **Comparison of adaptive pulse compression performance using SHG, LD, and LED feedback**. Results obtained for initial pulse durations of approximately 600 fs (a–c) and 260 fs (d–f). (a, d) Compressed pulse durations, (b, e) relative feedback signal gains, and (c, f) generations at which the maximum feedback signals were recorded. Individual markers represent four independent optimization runs, while horizontal lines and error bars indicate the mean values and standard deviations, respectively.

The feedback signal increase was quantified as the ratio of the maximum signal recorded during the optimization to the signal measured in the first generation. For pulses initially broadened to 600 fs, the mean feedback gains were 1.74 for SHG, 1.56 for LD, and 1.94 for LED (Fig. 4b). For initially 260 fs pulses, the corresponding values were 1.17, 1.16, and 1.19, respectively (Fig. 4e). These results show that all three feedback methods provided measurable signal enhancement, with larger relative increases for pulses requiring stronger compression.

The generation at which the maximum feedback signals were recorded was also compared across three detection methods (Fig. 4c, f). Despite variations between individual runs, the results indicate that LD, LED, and SHG feedback exhibit comparable behavior during optimization. The observed differences may be related to the stochastic nature of the optimization algorithms rather than the detection method themselves.

## 4. Mathematical model

To better understand the nature of the nonlinear detection process providing the feedback signal for pulse compression, we will now develop a simplified mathematical model. A TPA detector responds instantaneously to a time-varying optical power $P(t)$ equal to an intensity of the electric field scaled by the cross-sectional area. Grouping all relevant constants into $K_{\mathrm{TPA}}$ we obtain the nonlinear photocurrent:

$$I_{\mathrm{ph}} = K_{\mathrm{TPA}} P^2(t) \,. \tag{1}$$

After low-pass filtering (temporal averaging denoted by $\langle \cdots \rangle$), the mean value of the TPA photocurrent of a series of optical pulses with repetition rate $f_{\mathrm{rep}}$ can be expressed as:

$$\langle I_{\mathrm{ph}} \rangle = f_{\mathrm{rep}} K_{\mathrm{TPA}} \int_{-\infty}^{+\infty} P^2(t)\, \mathrm{d}t \,. \tag{2}$$

Our objective is now to relate the integral of the squared intensity to pulse duration in the constraint of a finite optical power (pulse energy).

### *4.1 Nonlinear photodetector as a pulse duration sensor*

For simplified treatment we will consider a Gaussian pulse with instantaneous power:

$$P_{\mathrm{G}}(t) = P_0 \exp\left[-4\ln(2)\left(\frac{t}{\tau}\right)^2\right], \tag{3}$$

where $P_0$ is the peak power and $\tau$ is the FWHM pulse duration. For a fixed repetition rate and neglecting material and geometrical constants, we obtain the following dependence:

$$\langle I_{\mathrm{ph}} \rangle \propto \int_{-\infty}^{+\infty} P_{\mathrm{G}}^2(t)\, \mathrm{d}t = P_0^2 \int_{-\infty}^{+\infty} \exp\left[-\frac{8\ln(2)}{\tau^2} t^2\right] \mathrm{d}t = P_0^2 \tau \sqrt{\frac{\pi}{8\ln(2)}}. \tag{4}$$

Now, we incorporate the physical constraints that the pulse energy $E_{\mathrm{p}}$ depends on the average power $\langle P_G(t) \rangle$. This means that there is a certain maximum peak power $P_0$ obtainable only for the Fourier-limited pulse – the shortest pulse possible for a given optical bandwidth. Recall that the energy of a Gaussian pulse for a given duration is

$$E_{\mathrm{p}} = P_0 \int_{-\infty}^{+\infty} \exp\left[-4\ln(2)\left(\frac{t}{\tau}\right)^2\right] \mathrm{d}t = P_0 \tau \sqrt{\frac{\pi}{4\ln(2)}} \approx 1.06 P_0 \tau \,. \tag{5}$$

After rearrangement, this equation establishes that the peak power for a given average power $P_{\mathrm{av}}$ is:

$$P_0 = \frac{E_{\mathrm{p}}}{\tau} \sqrt{\frac{4\ln(2)}{\pi}} = \frac{\langle P(t) \rangle}{f_{\mathrm{rep}} \tau} \sqrt{\frac{4\ln(2)}{\pi}} = \frac{P_{\mathrm{av}}}{f_{\mathrm{rep}} \tau} \sqrt{\frac{4\ln(2)}{\pi}} \,. \tag{6}$$

Now if we plug Eq. 6 into Eq. 4, $\langle I_{\mathrm{ph}} \rangle$ becomes expressed in terms of the average power and pulse duration.

$$\langle I_{\mathrm{ph}} \rangle \propto \frac{P_{\mathrm{av}}^2}{f_{\mathrm{rep}} \tau} \sqrt{\frac{2\ln(2)}{\pi}} \propto \frac{P_{\mathrm{av}}^2}{f_{\mathrm{rep}} \tau} \,. \tag{7}$$

Since for a given spectral width $\Delta\nu$ (FWHM) the pulse width has a lower bound given by the Fourier limit (analogue of the uncertainty principle $\Delta\nu\Delta\tau = 2\ln(2)/\pi$), there exists a global maximum of the nonlinear photocurrent when only the pulse duration varies (at a fixed average power). This makes the TPA detector useful for pulse duration assessment.

### *4.2 The case of higher order dispersion – multiple sub-pulses*

Higher order dispersion (HOD) effects, or the formation of bound soliton states, may lead to additional intensity peaks in the time domain. While the average power often remains almost constant, the time-averaged nonlinear photocurrent changes more drastically. The question that arises is whether those additional intensity peaks in the temporal profile lead to a non-uniqueness of the nonlinear photocurrent maximum since they also produce a TPA response. To study this effect, let us consider the extreme case of a Fourier-limited pulse breaking up into two pulses separated by *t'* with peak intensities relative to the single one given by *k* and 1–*k*, such that $0 < k < 1$. This analysis can be generalized to an arbitrary number of pulses; however the presented discussion is intended to provide the reader with physical intuition, which becomes less apparent when dealing with the analytical form of the temporal intensity profile for third-order dispersion in the form of a decaying pulse sequence modulated by a shifted and stretched Airy function [28].

Since the average nonlinear photocurrent is governed by the improper integral of a squared intensity, we aim to compare the integral of a clean single-peaked pulse

$$\int_{-\infty}^{+\infty} P^2(t)\,\mathrm{d}t \tag{8}$$

with that of the multi-peaked pulse. Our two-pulse model assumes that

$$P_2(t) = k\cdot P(t) + (1-k)\cdot P(t-t')\,. \tag{9}$$

Squaring $P_2(t)$ for integration yields

$$\left(P_2(t)\right)^2 = k^2P^2(t) + 2k(1-k)P(t)P(t-t') + (1-k)^2P^2(t-t')\,. \tag{10}$$

Assuming the integrals converge, we integrate both sides:

$$\int_{-\infty}^{+\infty}\left(P_2(t)\right)^2\mathrm{d}t = k^2\int_{-\infty}^{+\infty}P^2(t)\,\mathrm{d}t + 2k(1-k)\int_{-\infty}^{+\infty}P(t)P(t-t')\,\mathrm{d}t + (1-k)^2\int_{-\infty}^{+\infty}P^2(t-t')\,\mathrm{d}t\,. \tag{11}$$

Now we realize that one can express all the integrals in terms of $\int_{-\infty}^{+\infty}P^2(t)\,\mathrm{d}t$ because time shift $t\!-\!t'$ matters little here. Also, the second term of the sum resembles an overlap integral (here without considering interference), which is simply the improper integral of $P^2(t)$ scaled $0 < \eta < 1$ times. After these substitutions we arrive at:

$$\int_{-\infty}^{+\infty}\left(P_2(t)\right)^2\mathrm{d}t = \left[1-2k(1-k)(1-\eta)\right]\int_{-\infty}^{+\infty}P^2(t)\,\mathrm{d}t\,. \tag{12}$$

Now by studying the term in the square brackets we realize that the nonlinear photocurrent of the dual-peaked pulse never exceeds that of the single-peaked one. The term in the brackets can take values between 1/2 and 1 with the minimal value for $k = 1/2$ (two equal intensity pulses) and the maximal value for $k = 0$, $k = 1$, or $\eta = 1$ (perfect overlap equivalent to a single pulse). This agrees with intuition because the case of two equal-intensity pulses (analogous to a soliton doublet or diatomic soliton molecule) produces a $(1/2)^2$ nonlinear signal for each pulse relative to a single pulse, amounting to $2\times1/4=1/2$ of the signal produced by a single pulse in total.

From this analysis we see that a maximum of the nonlinear photocurrent corresponds to the case when all optical power is localized in a single pulse. A multi-peaked pulse with the same FWHM duration always produces a lower time-averaged TPA signal if the average power is the same. This proves the existence of the global maximum required for the feedback system.

## 5. Discussion and conclusions

The use of commercially available AlGaAs diodes as nonlinear photodetectors offers an interesting alternative to conventional SHG-based feedback. While SHG relies on frequency conversion in a nonlinear crystal followed by optical detection, semiconductor devices generate an electrical signal directly through nonlinear absorption. The observed dependence on quadratic average power and inverse on pulse duration are consistent with the analytical model. The model further predicts that, for a fixed average power and spectral bandwidth, the nonlinear response reaches its maximum for the shortest, single-peaked pulse rather than one split into temporally separated components. This provides a clear basis for using photocurrent maximization as the objective for adaptive pulse compression.

The practical usefulness of semiconductor-based feedback depends not only on the magnitude of the generated photocurrent but also on the optical power range over which a useful nonlinear response can be maintained. Importantly, both devices exhibited a measurable nonlinear response even at average optical powers below 1 mW. The higher photocurrent generated by the LD may be associated with its waveguide geometry and optical coupling conditions. However, its deviation from quadratic behavior at higher optical powers suggests a more restricted operating range than that observed for the LED. Saturation effects have previously been reported in semiconductor light-emitting devices at incident optical power of approximately 40 mW [22], where photovoltage saturation was attributed to the electronic properties of the p–n junction, suggesting a possible explanation for the LD response observed here. The absence of observable saturation in the LED over the investigated power range makes it particularly promising for applications requiring greater flexibility in incident optical power.

The comparison with SHG also provides insight into the practical performance of the investigated detection methods. Although the semiconductor devices generated stronger electrical signals in the present setup, the measured amplitudes also depend on the detection arrangement. In particular, the SHG reference employed a silicon photodetector, whereas alternative schemes using photomultiplier tubes could offer higher sensitivity to weak SHG signals [8, 29]. Beyond signal magnitude, the dependence on incident polarization is another important consideration for practical feedback detection. The weaker polarization dependence of the LD and LED compared with the SHG reference may reduce the need for precise polarization control in optical systems. Importantly, despite these differences in detector characteristics, both semiconductor devices enabled pulse compression to durations comparable to those obtained using SHG feedback. The LED achieved particularly similar results under both initial dispersion conditions, demonstrating that a stronger photocurrent does not necessarily lead to shorter compressed pulses. The similar overall profiles of the optimized SLM voltage masks further suggest that all three feedback methods guided the optimization toward comparable spectral phase corrections.

One of the main advantages of semiconductor-based feedback is the simplicity of the detection arrangement. Unlike conventional SHG-based systems, the investigated diodes do not require a nonlinear crystal, spectral filtering of the fundamental radiation, or a separate optical photodetector, while also allowing operation under normal laboratory lighting conditions. Their compact size and commercial availability may further facilitate integration into existing laser systems. However, this simplicity comes with certain limitations in the information provided by the feedback signal. While semiconductor detectors provide a scalar signal related to the nonlinear optical response, techniques based on spectrally resolved SHG, such as MIIPS [30], can provide additional information for spectral phase characterization and correction. Therefore, the choice of feedback method depends on the intended application: semiconductor devices offer a straightforward approach for automated pulse optimization, whereas spectrally resolved methods may be preferable when more detailed phase characterization is required.

Overall, the results demonstrate that commercially available AlGaAs LDs and LEDs can provide effective nonlinear feedback for adaptive femtosecond pulse compression, achieving performance comparable to conventional SHG-based detection with a simpler optical arrangement. By selecting semiconductor devices with suitable bandgap energies, this approach could potentially be extended to other wavelength ranges. The compact size and reduced optical complexity of semiconductor-based detectors may facilitate their integration into existing ultrafast laser systems, enabling more convenient and automated dispersion compensation in applications such as multiphoton microscopy.

**Funding.** Fundacja na rzecz Nauki Polskiej (FENG.02.02-IP.05-0121/25); Narodowe Centrum Nauki (2022/45/B/ST7/03316).

**Acknowledgment.** The project "Two-photon fluorescence lifetime ophthalmoscopy: new perspectives in the diagnosis of eye diseases" (FENG.02.02-IP.05-0121/25) is carried out within the First Team programme of the Foundation for Polish Science co-financed by the European Union under the European Funds for Smart Economy 2021-2027 (FENG). This work is supported by the use of the National Laboratory for Photonics and Quantum Technologies (NPLQT) infrastructure, which is financed by the European Funds under the Smart Growth Operational Programme.

**Disclosures.** The authors declare no conflicts of interest.

**Data availability.** Data underlying the results presented in this paper are not publicly available at this time but may be obtained from the authors upon reasonable request.